# Identificación de once candidatos vacunales potenciales contra la malaria por medio de la bioinformática

## Identification of 11 potential malaria vaccine candidates using Bioinformatics

Raúl Isea*

Fundación Instituto de Estudios Avanzados IDEA. Valle de Sartenejas, Hoyo de la Puerta, Baruta 1080, Venezuela.

e-mail: risea@idea.gob.ve

En este trabajo se propusieron once candidatos vacunales potenciales contra el *Plasmodium falciparum*, causante de la muerte de entre 2 y 3 millones de personas cada año. Los candidatos vacunales se seleccionaron mediante una búsqueda de fragmentos del genoma de *P. falciparum* que presentaban una clara evidencia antigénica y que, a su vez, están expresadas en cuatro estadios del ciclo de vida del parásito: esporozoito, merozoito, trofozoito y gametocito. Los candidatos finales, elegidos *in silico* con ayuda de las herramientas de Bioinformática, son aquellos fragmentos ya secuenciados y contenidos en la base de datos PlasmoDB. Los identificadores de los candidatos son: PFC0975c, PFE0660c, PF08_0071, PF10_0084, PFI0180w, MAL13P1.56, PF14_0192, PF13_0141, PF14_0425, PF13_0322 y PF14_0598.

**Palabras clave**: Mapeo de epitopos, *Plasmodium falciparum,* malaria, bioinformática.

In this paper, we suggested eleven protein targets to be used as possible vaccines against Plasmodium falciparum causative agent of almost two to three million deaths per year. A comprehensive analysis of protein target have been selected from the small experimental fragment of antigen in the P. falciparum genome, all of them common to the four stages of the parasite life cycle (i.e., sporozoites, merozoites, trophozoites and gametocytes). The potential vaccine candidates should be analyzed in silico technique using various bioinformatics tools. Finally, the possible protein target according to PlasmoDB gene ID are PFC0975c, PFE0660c, PF08_0071, PF10_0084, PFI0180w, MAL13P1.56, PF14_0192, PF13_0141, PF14_0425, PF13_0322, y PF14_0598.

**Introducción**

La malaria o paludismo es una enfermedad tropical, causada por un parásito del género *Plasmodium*, a través de la picadura de un mosquito infectado (*Anopheles gambiae)*, que causa entre 2 y 3 millones de muertes anualmente y la mitad son niños menores de cinco años. Hasta ahora se conocen cuatro cepas diferentes que afectan a la especie humana: *P. falciparum*, *P. vivax*, *P. ovale y P. malariae*. De todas ellas, *P. falciparum* es la más virulenta, de 500 a 800 millones de casos anuales en el mundo. Según la Organización Panamericana de la Salud (OPS), durante el año 2007, la malaria afectó a 21 países, con una población total de unos 880 millones de personas. De ellas, 236 millones habitan en zonas endémicas, mientras que otros 276 millones viven en áreas de riesgo de transmisión (1). En el caso de la República Bolivariana de Venezuela, el riesgo de contagio afectó el 29% de la población, según datos suministrados por la OPS (1). Además, hay que tener en cuenta que de los 916.000 km$^2$ que constituye la superficie total de Venezuela, 430.000 km$^2$ son terreno selvático, ideal para la propagación de mosquitos contaminados.

En 1998, la Organización Mundial de la Salud, bajo el lema *Hacer retroceder la malaria* para el 2015 (2), auspició una campaña a escala mundial para avanzar en su erradicación. La misma fue financiada con un fondo de 3.000 millones de dólares que fueron insuficientes, si lo comparamos con los gastos médicos y pérdidas económicas que dicha enfermedad causa sólo en África (unos 30.000 millones de dólares). En este sentido, la Organización de las Naciones Unidas ha manifestado que se requirirían otros 3.000 millones de dólares para el 2010 para cumplir los objetivos del plan de erradicación propuesto hace una década (3).

---

* Dr. Raúl Isea. Instituto de Estudios Avanzados-IDEA. Hoyo de la Puerta, Baruta, Venezuela

Las esperanzas de combatir dicho flagelo están puestas en la vacuna, conocida con el nombre RTS,S/AS02, elegida entre un total de 17 que son ensayadas actualmente. Ésta ha demostrado ser segura y, a su vez, produjo una disminución del número de episodios clínicos en un 30%, según datos del ensayo realizado en Mozambique en 2.022 voluntarios, con edades comprendidas entre 1 a 4 años, con un período de protección que asciende a 45 meses (4).

La producción a escala masiva se encuentra en su última fase, gracias al esfuerzo mancomunado del Institut de Recherche en Science de la Santé, Kumasi Centre for Collaborative Research & School of Medical Sciences Kumasi, Kintampo Health Research Centre, Kenya Medical Research Institute, KEMRI Wellcome Collaborative Research Programme, University of North Carolina Project, Centro de Investigação em Saude de Manhiça, Ifakara Health Research Development Centre, National Institute for Medical Research, Albert Schweitzer Hospital - MRU, y GlaxoSmithKline; quienes recibieron también apoyo económico de la Fundación Bill Gates a través de Path Malaria Vaccine Initiative.

Esta vacuna posee dos componentes denominados RTS y S, que se obtuvieron por recombinación del gen que codifica para la proteína circunsporozoítica del *P. falciparum*, derivada de una sección de la región C-terminal, a través de la cual se une al extremo amínico del antígeno de superficie del virus de la hepatitis B (4). El coadyuvante que se empleó fue denominado AS02A y consiste en una mezcla de aceite con agua, donde está el monofosforil lípido A (conocido por sus siglas en inglés, MPL) más un derivado no tóxico de la saponina purificada por HPLC, obtenida de la corteza del árbol *Quillaja Saponaria* Molina, denominado QS21. Dicho coadyuvante ayuda a desencadenar la respuesta celular de los linfocitos T (4).

Sin embargo, mientras se pueda obtener una vacuna, hay que continuar desarrollando otras alternativas más eficientes que permitan, entre otras cosas, aumentar el período de protección por más de 45 meses.

Las drogas antimaláricas son medicamentos básicamente derivados de la quinolona (*ie.*, cloroquina, quinina, y mefloquina), los cuales parecen actuar en la ruta de degradación de la hemoglobina al unirse con la ferriprotoporfirina IX (5); los compuestos derivados de cloroquina, que parecen actuar destoxificando el grupo hemo; los antifolatos, como son las sulfamidas, que inhiben la acción de la dihidropteroato sintasa (del inglés, *Dihydropteroate synthetase*), y por último, la pirimetamina que bloquea una enzima responsable de la regeneración del cofactor en estado reducido. Dichos tratamientos no han sido eficaces, porque las proteínas blanco donde actúan estas drogas han generado resistencia, así como hipersensibilidad a dichos medicamentos durante su tratamiento profiláctico (6).

En paralelo, gracias a los avances que aporta la biología molecular a través de la genómica, así como los datos experimentales generados por la proteómica, se han logrado identificar aquellas proteínas que se expresan en cuatro estadios del parásito, es decir, el esporozoito, merozoito, trofozoito y gametocito (7).

Por ello, el objetivo de este trabajo fue la selección de candidatos vacunales potenciales a partir de una búsqueda completa en todo el genoma *P. falciparum* (8). Bajo el criterio de que deben ser antigénicos y expresados en los cuatro estadios del ciclo de vida del parásito. Esta última condición impuesta fue con el objetivo de obtener una mayor efectividad, a pesar de que no existe todavía ninguna evidencia experimental de su necesidad.

**Metodología**

En primer lugar, se identificaron aquellas secuencias antigénicas que están anotadas en el genoma del *P. falciparum* (8), con especial atención en aquellas proteínas que puedan interferir en la ruta metabólica del parásito, pero que a su vez no esté presente en el huésped. Este último requisito intenta garantizar la no toxicidad en el anfitrión, aunque dicha afirmación no esté aún fundamentada con evidencia experimental. Con vistas a que dicho genoma posee muchas secuencias anotadas como hipotéticas se detectó el carácter antigénico empleando los algoritmos publicados por Kolaskar y Tongaonkar (10), así como los predichos por accesibilidad (11). Junto a esto se realizó también una búsqueda bibliográfica de todos aquellos trabajos que están disponibles en PubMed (12), donde se manifiesta un carácter antigénico a partir de evidencia experimental. Dichas consultas se realizaron empleando las duplas: 'drug AND plasmodium', 'plasmodium AND antigenicity', entre otras combinaciones.

Por otra parte, se emplearon los resultados obtenidos del proteoma del *P. falciparum* (7), ya que nos permitió identificar los cambios en la expresión de las proteínas durante las distintas fases en el ciclo del vida del parásito, es decir, aquellas que se expresan en sus estadios: esporozoito, merozoito, trofozoito y gametocito.

Por último, se realizó la intersección de aquellas proteínas presentes en los cuatro estadios --indicados anteriormente-- con aquellos que presentaron evidencia antigénica experimental. Dicha selección se validó a partir de la búsqueda de un consenso de similitud con otras secuencias, empleando para ello el programa mpiBlast, según la metodología descrita en el trabajo que se publicó el año pasado (13). Más aún, se verificó dicha similitud con ayuda de las bases de datos OrthoMCL (14), que permitió identificar secuencias ortólogas, entre la gama posible de resultados obtenidos por el programa mpiBlast (13). Dichos candidatos vacunales debieron estar ubicados en la región externa de la membrana celular, y para ello se empleó el programa TMHMM2 que permitió predecirlos (15).

**Resultados y Discusión**

El genoma de la cepa *Plasmodium falciparum* 3D7 (taxonomía ID 36329) fue secuenciado en 2002 (8). El mismo es de 23,26 Mb y está compuesto por 14 cromosomas lineales, unidos en cada extremo a secuencias teloméricas, cuyos cromosomas oscilan entre 0,75 Mb (154 genes en el cromosoma 1) y 3,5 Mb (771 genes en el cromosoma 14). Esta información está disponible a través de la base de datos PlasmoDB (16), de acceso gratuito a través de la página web: http://plasmodb.org. En dicha base de datos existen 5.373 genes, 10.999 proteínas, 78 seudogenes y 73 genes ARN.

En paralelo, el resultado obtenido de la búsqueda bibliográfica de todos aquellos trabajos en los que se manifestara un carácter antigénico generó una lista de 22.543 entradas, que una vez depurada condujo a una lista de 515 potenciales proteínas candidatas (actualmente se esta diseñando una base de datos con toda esta información a la que se puede acceder desde Internet). Sin embargo, la principal limitación de este método es que la mitad de las proteínas en dicho genoma son hipotéticas, y por tanto no se pudieron tener en cuenta en el presente estudio. Para evitar este sesgo en la información es necesario determinar las propiedades antigénicas de todas las regiones de unión al complejo mayor de histocompatibilidad de las proteínas que están presentes en el genoma de *P. falciparum*. Sin embargo, el alto número de falsos positivos generados por los programas de predicción es muy elevado, y curiosamente no coincide con los datos obtenidos por los métodos experimentales.

A modo de ejemplo, se muestra en la Tabla 1 la predicción de epítopos de todas aquellas secuencias que están presentes en el cromosoma 1 del genoma de *P. falciparum*, formado por 643.292 nucleótidos, de los cuales 143 son proteínas, según la metodología descrita por Kolaskar y Tongaonkar (10), así como los predichos por accesibilidad (11). La intersección de esta información con la que se obtuvo de la literatura científica, en donde se han empleado secuencias antigénicas, nos permitió agrupar todos los epítopos correspondientes a cada cromosoma. En la Tabla 1 se muestran aquellos epítopos generados de acuerdo con los métodos de predicción (10-11), así como los publicados en revistas científicas.

**Tabla 1.** Epítopos predichos por el método de accesibilidad (11), por Kolaskar y Tongaonkar (10).

| Predichos por accesibilidad | Predichos por Kolaskar y Tongaonkar | Citados en la literatura científica |
|---|---|---|
| KNVKEKNP | FHAYSWIFSQ | DDEHVEEPTVA |
| FDDEEKRNEN | FLKVLCSKRGVLPIIGILYIILN | DDEHVEEPTVADDEHVEEPTVA |
| VNRYRYSNN | SSGVQFTDR | DVVGYIMHGISTINKEMK |
| INKRKYDSL | GETLPVNPY | EENVEENV |
| EKLQKTYSQYKV | NPIVVSQVFGLPFEK | EENVEENVEENV |
| DMPKEAYESK | TFTLESP | EENVEENVEENVEENV |
| KNWYRQK | AIPHISEFNPLIVDKVLFD | EENVEENVEENVEENVEENV |
| EMKKRAQKPKKKKSR | QKTYSQYKVQYD | EENVEHDA |
| VEPQQEEP | WTQCIKLI | EENVEHDAEENVEENV |
| LEKQKY | QKYLNLE | EENVEHDAEENVEHDA |
| ESVKENEEEH | RLTVLNQI | EENVEHDAEENVEHDAEENVEENV |
| NYYPYQRS | SNQIQYSCR | MQTLWDEIMDINKRK |
| IDKKRWYNKYG | GWLCCGG | SLRWIFKHVAKTHLK |
| QYQKER | EEPVQTVQEQ | TVAEEHVEEPTVAEE |
| HLKKSSKSAKKLQQRTQANKQE | DILPSLRAS | EENV |
| KTLKKRAQS | INYYDT | |
| KRNDKKAKKYDT | KDGVYLDH | |
| KEMKNQNENV | DEDLLFDL | |
| NVEEYDEEN | EEPASDVQQT | |

En la Tabla 2 se agruparon los resultados de los epítopos detectados por métodos experimentales en los clusters, es decir, los epítopos son agrupados de acuerdo con su composición aminoacídica y clasificados según sus propiedades fisicoquímicas. Sorprende el pequeño número de clusters que están presentes en los cromosomas, pero hay que tener presente que en el proceso de análisis se descartaron aquellos epítopos que se repetían, como ocurrió en las secuencias identificadas como PFA0110w, PFA0140c, PFA0225w, PFA0280w, PFA0350c, PFA0625w, PFA0665w. La secuencia proteica NKND está presente en PFA0665w, PFA0625w, PFA0350c, y PFA0280w; el resto de los epítopos son: TDVNRYRYSNN, YEAIPHIS, EENVEHDA, EENVEENV, entre otros.

Al reexaminar los resultados que se muestran en la Tabla 2 se deduce que los anticuerpos dirigidos contra este antígeno que corresponde a la secuencia de aminoácidos repetitiva EENV (expresada a nivel de la superficie globular), es un factor importante en la producción de anticuerpos durante la infección de la malaria. Sin embargo, dicho patrón no pudo predecirse con los programas de Bioinformática y se requieren mayores esfuerzos en biología computacional para poder extender este resultado a otras cepas.

**Tabla 2.** Epítopos detectados por métodos experimentales agrupados en clusters

| Cl | Gen | Epítopos |
|---|---|---|
| 1 | PFA0110w | EENVEENV, EENVEHDA, EENVEENVEENVEENVEENVEENVEENVEENV |
| 2 | PFA0665w | IKND, NKND |
| 3 | PFA0635c | NHDTRINDYNKRLTEYNKRLTEYNKRLTEYTKRLNE |
| 4 | PFA0110w | TDVNRYRYSNNYEAIPHIS |

Cl representa el clúster que agrupan los epítopos obtenidos en las proteínas antigénicas, mientras que la segunda y tercera columna representan respectivamente el identificador del gen de acuerdo con la anotación de PlasmoDB (16) y los epítopos obtenidos en el cromosoma 1 del *P. falciparum*.

En la Tabla 3 se han recogido todas aquellas proteínas presentes en el genoma de *P. falciparum* que se expresan en cada uno de los cuatro estadios del parásito: esporozoito, merozoito, trofozoito y gametocito (7). En dicha Tabla solo se indica el identificador (ID) designado en la base de datos PlasmoDB (16).

El siguiente paso consistió en comparar los resultados obtenidos de las proteínas presentes tanto en la Tabla 2 y 3, de lo que se puede deducir aquellas proteínas que puedan ser consideradas como candidatos vacunales, las cuales muestran su carácter antigénico y a su vez están condicionadas a su expresión en los cuatro estadios del ciclo de vida del parásito. Más aún, en dichos candidatos se examino su similitud con otras secuencias derivadas de los resultados obtenidos de mpiBlast (13) y OrthoMCL (14) que deben estar ubicados en la región externa de la membrana, de acuerdo con los resultados derivados del programa TMHMM2 (15).

**Tabla 3.** Número de acceso de las proteínas antigénicas de acuerdo con la nomenclatura asignada en el PlasmoDB (16)

| ID | ID | ID | ID | ID | ID |
|---|---|---|---|---|---|
| MAL13P1.270 | MAL13P1.283 | MAL13P1.308 | PF08_0019 | MAL13P1.308 | PF10_0079 |
| MAL13P1.323 | MAL13P1.333 | PFA0400c | PF08_0063 | PFA0400c | PF10_0153 |
| MAL6P1.160 | MAL6P1.189 | MAL6P1.232 | PF11_0098 | MAL6P1.232 | PF11_0331 |
| MAL6P1.244 | MAL6P1.248 | MAL6P1.249 | PF11_0158 | MAL6P1.249 | PF13_0014 |
| MAL6P1.254 | MAL6P1.48 | MAL6P1.91 | PF11_0208 | MAL6P1.91 | PF13_0070 |
| MAL7P1.147 | MAL7P1.81 | MAL8P1.103 | PF11_0272 | MAL8P1.103 | PF13_0233 |
| MAL8P1.113 | MAL8P1.125 | MAL8P1.17 | PF11_0352 | MAL8P1.17 | PF13_0305 |
| MAL8P1.69 | MAL8P1.72 | MAL8P1.83 | PF13_0065 | MAL8P1.83 | PF13_0328 |
| PF13_0316 | PF14_0655 | PF14_0261 | PF13_0262 | PF10_0063 | PF11_0096 |
| PF14_0083 | PF14_0192 | PF14_0344 | PF14_0407 | PF10_0081 | PF11_0111 |
| PF14_0315 | PfM18AAP | PFB0260w | PF11_0183 | PFB0260w | PF08_0034 |
| PF14_0368 | PF14_0324 | PF07_0033 | PF11_0270 | PF07_0033 | PF10_0155 |
| PF14_0443 | PF14_0391 | PF07_0088 | PF11_0351 | PF07_0112 | PF10_0268 |
| PF14_0510 | PF14_0448 | PF14_0486 | PF13_0033 | PF08_0054 | PF11_0043 |
| PF14_0627 | PF14_0517 | PF14_0615 | PF13_0079 | PF08_0075 | PF11_0062 |
| PFC0275w | PFC0285c | PFC0400w | PF08_0113 | PFC0400w | PF10_0213 |
| PFC0350c | PFC0735w | PFC0920w | PF10_0068 | PFC0920w | PF10_0320 |
| PFC0900w | PFD0665c | PFD0095c | PF10_0115 | PFD0095c | PF11_0061 |
| PFD1060w | PFD0305c | PFE0585c | PF10_0210 | PFE0585c | PF11_0069 |
| PFE0165w | PFE0225w | PFE0255w | PF11_0065 | PFE0255w | PF11_0250 |
| PFE0810c | PFE0865c | PFE0870w | PF10_0272 | PFE0870w | PF11_0098 |
| PFE1120w | PFE1195w | PFE1370w | PF11_0055 | PFE1370w | PF11_0175 |
| PFI0645w | PFI0875w | PFI0930c | PF13_0179 | PFI0930c | PF14_0174 |
| PFI1105w | PFI1475w | PFI1525w | PF13_0304 | PFI1525w | PF14_0246 |
| PFL0590c | PFL0670c | PFL1070c | PFI1570c | PFL1070c | PF08_0034 |
| PFL1390w | PFL1425w | PFL1550w | PFB0445c | PFL1550w | PF08_0074 |
| PFL1725w | PFL2215w | PFL0310c | PF07_0054 | PFL0310c | PF10_0041 |

Tras aplicar el procedimiento descrito, se obtienen once candidatos vacunales potenciales que son antigénicas, se expresan en cuatro de los estadios del ciclo de vida del parásito, están ubicados en la parte externa de la membrana, y a su vez son similares con otras secuencias proteicas. Por todo ello, los candidatos obtenidos son aquellos cuyos identificadores en la base de datos PlasmoDB corresponden a (16): PFC0975c, PFE0660c, PF08_0071, PF10_0084, PFI0180w, MAL13P1.56, PF14_0192, PF13_0141, PF14_0425, PF13_0322, y PF14_0598 (Tabla 4).

Se debe destacar la importancia de un trabajo interdisciplinario que permita encontrar soluciones contra esta enfermedad que afectan gravemente a la salud pública. Este es el ejemplo que nos ofrecen los diversos centros que participan en la fase II de la vacuna experimental RTS, S/AS02. En este sentido, nos alegraría el fomento de iniciativas internacionales como es el caso de WISDOM (abreviatura de las siglas en inglés que significa: "*Wide In-Silico Docking Of Malaria*") para la búsqueda de nuevas drogas contra la malaria (9). Igualmente es de desear la formación y consolidación de redes internacionales de trabajo y a modo de ejemplo, mencionar la Red Latinoamericana de Información Científico Técnico en Vacunas, financiada por el Grupo 77 de la Naciones Unidas, la Red Suramericana e Iberoamericana de Bioinformática, auspiciado por CNPq en Brasil, la Red Iberoamericana de Tecnologías Convergentes NBIC en Salud con aportes de Cyted, el Instituto Intercientífico de Paleopatología y Derechos Genoculturales, entre otros.

**Tabla 4.** Número de acceso del candidato vacunal de acuerdo con la nomenclatura en PlasmoDB (16), localización en el genoma e identificación en el cromosoma. Identificación del grupo ortológico de acuerdo con la base de datos OrthoMCL (14), y los respectivos alias en su nomenclatura

| ID | Localización | Cromosoma | Grupo Ortológico | Alias |
|---|---|---|---|---|
| PFC0975c | 925315 .. 925830 | 3 | OG1.2_93 | CPR3, PfCyP19, MAL3P7.25 |
| PFE0660c | 569180 .. 569917 | 5 | OG1.2_4961 | MAL5P1.133 |
| PF08_0071 | 709349 .. 709945 | 8 | OG1.2_236 | FeSOD, PfSOD1 |
| PF10_0084 | 360625 .. 362480 | 10 | OG1.2_138 | |
| PFI0180w | 175540 .. 177440 | 13 | OG1.2_153 | |
| MAL13P1.56 | 501992 .. 505249 | 13 | OG1.2_5282 | |
| PF14_0192 | 822418 .. 824197 | 14 | OG1.2_190 | |
| PF13_0141 | 1040792 .. 1042657 | 13 | OG1.2_359 | PfLDH |
| PF14_0425 | 1843984 .. 1845894 | 14 | OG1.2_436 | |
| PF13_0322 | 2431676 .. 2435257 | 13 | OG1.2_1397 | flN |
| PF14_0598 | 2558487 .. 2560920 | 14 | OG1.2_104 | GAPDH |

Los ID de cada uno de los candidatos vacunales son: PFC0975c = peptidil prolil cis-trans isomerasa (del inglés, *peptidyl-prolyl cis-trans isomerase*); PFE0660c = Purina-nucleosido Fosforilasa (del inglés, *purine nucleotide phosphorylase*); PF10_0084 = cadena beta tubulina (del inglés, *tubulin beta chain*); PFI0180w = alfa tubulina (del inglés, *alpha tubulin*); MAL13P1.56 = familia m1 de las aminopeptidasa (del inglés, *m1-family aminopeptidase*); PF14_0192 = glutatión reductasa (del inglés, *glutathione reductase*); PF13_0141 = L-deshidrogenasa láctica (del inglés, *L-lactate dehydrogenase*); PF14_0425 = fructosa-bisfosfato aldolasa (del inglés, *fructose-bisphosphate aldolase*); PF13_0322 = falcilisina (del inglés, *falcilysin*); PF14_0598 = Gliceraldehído-3-fosfato deshidrogenasa (del inglés, *glyceraldehyde-3-phosphate dehydrogenase*).

## Conclusiones

El presente trabajo ha puesto de manifiesto la necesidad de emplear datos obtenidos de los trabajos experimentales a la hora de seleccionar candidatos vacunales potenciales contra el *P. falciparum*. De ellos, se ha extraído una lista 515 proteínas posibles. Cuando se añade la condición de que dichos candidatos deben expresarse en los cuatros estadios del parásito, y a su vez debe presentar similitud funcional con otras secuencias ortólogas obtenidas con la ayuda de la base de datos OrthoMCL, el número se reduce drásticamente a las once seleccionadas. Sólo resta confirmar la aplicabilidad de este análisis a través de la implementación de métodos experimentales, como por ejemplo, amplificar las secuencias a través del uso de PCR, para que posteriormente sean clonadas, expresadas y purificadas como proteína recombinante, para su posterior ensayo experimental, el cual requiere de una batería de métodos ómicos que permitan comprender el papel de cada uno de los candidatos vacunales encontrados.

## Referencias

1. Organización Panamericana de la Salud. Population Living in Malaria-Endemic Areas in the Americas 1994–2007. Disponible en: http://www.paho.org/English/AD/DPC/CD/mal-americas-2007.pdf . [Consultado: 19 June 2008].

2. Nabarro DN, Tayler EM. The roll back malaria campaign. Science 1998;280:2067-8.

3. UNICEF. El Paludismo. Disponible en: http://www.unicef.org/spanish/health/index_malaria.html [Consultado: 29 September 2009].

4. Sacarlal J, Aide P, Aponte JJ, Renom M, Leach A, Mandomando I, et al. Long-Term Safety and Efficacy of the RTS,S/AS02A. Malaria Vaccine in Mozambican Children. J Infect Dis 2009; 200:329-36.

5. Loria P, Miller S, Foley M, Tilley L. Inhibition of the peroxidative degradation of haem as the basis of action of chloroquine and other quinoline antimalarials. Biochem J 1999; 339: 363-70.

6. Astro-Sancho J, Munguia-Ramírez M, Avila-Agüero M. Malaria: una actualización. Acta Méd Costarric 2002; 44:107-112.

7. Lal K, Prieto JH, Bromley E, Sanderson SJ, Yates JR 3rd, Wastling JM, et al. Large-Scale Differential Proteome Analysis in Plasmodium falciparum under Drug Treatment. PLoS ONE 2008; 3: e4098.

8. Gardner MJ, Hall N, Fung E, White O, Berriman M, Hyman RW, et al. Genome sequence of the human malaria parasite Plasmodium falciparum. Nature 2002; 419: 498-511.

9. Kasam V, Salzemann J, Botha M, Dacosta A, Degliesposti G, Isea R, et al. WISDOM-II: Screening against multiple targets implicated in malaria using computational grid infrastructures. Malaria J 2009; 8:88.

10. Kolaskar AS, Tongaonkar P. A semi-empirical method for prediction of antigenic determinants on protein antigens, FEBS Lett 1990; 276:172-4.

11. Emini EA, Hughes JV, Perlow DS, Boger J. Induction of hepatitis A virus-neutralizing antibody by a virus-specific synthetic peptide. J Virol. 1985; 55:836-9.

12. Wheeler DL, Barrett T, Benson DA, Bryant SH, Canese K, Chetvernin V, et al. Database resources of the National Center for Biotechnology Information. Nucleic Acids Res. 2006; 34:D173-D180.

13. Aparicio G, Blanco F, Blanquer I, Bonavides C, Chaves JL, Embid M, et al. Developing Biomedical Applications in the framework of EELA. En: Udoh E, Wang FZ, eds. Handbook of Research on Grid Technologies and Utility Computing: Concepts for Managing Large-Scale Applications. Hersey, EE.UU: Editorial IGI Global; 2009. p. 206-18.

14. Chiu JC, Lee EK, Egan MG, Sarkar IN, Coruzzi GM, DeSalle R. Ortholog ID automation of genome-scale ortholog identification within a parsimony framework. Bioinformatics 2006; 22: 699-707.

15. Moller S, Croning MDR, Apweiler R. Evaluation of methods for the prediction of membrane spanning regions. Bioinformatics 2001;17:646-53.

16. Aurrecoechea C, Brestelli J, Brunk BP, Dommer J, Fischer S, Gajria B, et al. PlasmoDB: a functional genomic database for malaria parasites. Nucleic Acids Res. 2009;37:D539-D43.